\documentclass[a4paper]{article}

\usepackage{17lomcon}        
\usepackage{cite}             
\usepackage{epsfig}           

\begin{document}


\title{Jiangmen Underground Neutrino Observatory: Status and Prospectives
\footnote{Talk presented at the 17th Lomonosov Conference on Elementary Particle Physics, 20-26 August 2015, Moscow State University.}}

\author{Yu-Feng Li \email{liyufeng@ihep.ac.cn}\\
(On behalf of the JUNO collaboration)
}

\affiliation{Institute of High Energy Physics, Chinese Academy of Sciences,
P.O. Box 918, Beijing 100049, China}


 \maketitle


\begin{abstract}
The Jiangmen Underground Neutrino Observatory (JUNO) is a 20 kton liquid scintillator (LS) detector,
which is planed to determine the neutrino mass hierarchy and measure the oscillation parameters
at the sub-percent level using reactor antineutrino oscillations. As a multipurpose neutrino experiment,
JUNO is also capable of measuring supernova burst neutrinos, the diffuse supernova neutrino background, geo-neutrinos,
solar neutrinos and atmospheric neutrinos. After a brief introduction to the physics motivation,
we discuss the status of the JUNO project, including the design of the detector systems.
Finally the latest civil progress and future prospectives are also highlighted.
\end{abstract}

\section{Introduction}
The recent discovery of non-zero $\theta_{13}$ in reactor~\cite{DYB,DC,RENO} and accelerator~\cite{T2K,MINOS} neutrino experiments has opened a new window
for neutrino oscillation studies. With relatively large $\theta_{13}$, the unknown neutrino mass hierarchy (i.e., the sign of $\Delta m^2_{31}$ or $\Delta m^2_{32}$)
and lepton CP violation are anticipated to be measured in the near future.
The determination of the neutrino mass hierarchy (MH) is applicable in the long-baseline accelerator neutrino experiments~\cite{HK,DUNE,MOMENT},
atmospheric neutrino experiments~\cite{INO,PINGU}, and medium baseline reactor antineutrino experiments \cite{JUNO,JUNO2,RENO50}. 

\begin{figure}
\begin{center}
\begin{tabular}{c}
\includegraphics*[width=0.7\textwidth]{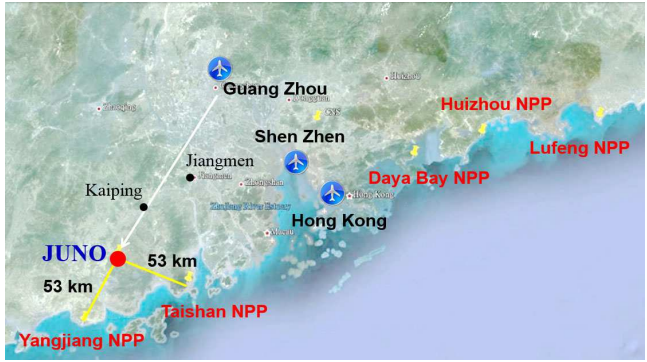}
\end{tabular}
\end{center}
\caption{Layout of the experimental design for JUNO. The Yangjiang and Taishan nuclear reactor
complexes are used for reactor antineutrino studies. \label{location}}
\end{figure}

The Jiangmen Underground Neutrino Observatory (JUNO) is a multipurpose liquid scintillator (LS) neutrino experiment, whose
primary goal is to determine the neutrino mass hierarchy and measure the oscillation parameters
at the sub-percent level using reactor antineutrino oscillations~\cite{JUNO,sens}. The layout of the JUNO experiment~\cite{JUNO2} is shown
in Fig.~\ref{location}, where the experimental site is located at Jiangmen in South China, $\sim$53 km away from the Yangjiang and Taishan reactor
complexes, which are planed to have six reactor cores of 2.9 GW$_{th}$ and four reactor cores of 4.6 GW$_{th}$, respectively.
JUNO will build a LS detector of 20 kton with the 3\%/$\sqrt{E}$ energy resolution.
The detector is planed to be located in an experimental hall with the overburden of larger than 700 meters for the purpose of reducing the muon-induced backgrounds.

\section{Physics motivations}
\label{sec:physics}

The neutrino MH is among the most important issues in the future neutrino oscillation program.
It can be resolved in reactor antineutrino experiments at the medium baseline
using interference between the two fast oscillation components. 
Because the oscillation amplitudes of these fast components are different (i.e., $\sin^2\theta_{12}<\cos^2\theta_{12}$)
and the relative size of their frequencies is also different for two neutrino MHs ($|\Delta m^2_{31}|>|\Delta m^2_{32}|$ or $|\Delta
m^2_{31}|<|\Delta m^2_{32}|$), the interference between the fast oscillation components provides us the discrimination ability
of the MH. 
Additional information of the MH can be obtained by comparing the effective atmospheric mass-squared differences in
experiments of the medium baseline reactor antineutrino oscillation and long baseline accelerator muon (anti)neutrino disappearance, respectively.

\begin{figure}
\begin{center}
\begin{tabular}{cc}
\includegraphics*[bb=25 20 295 228, width=0.46\textwidth]{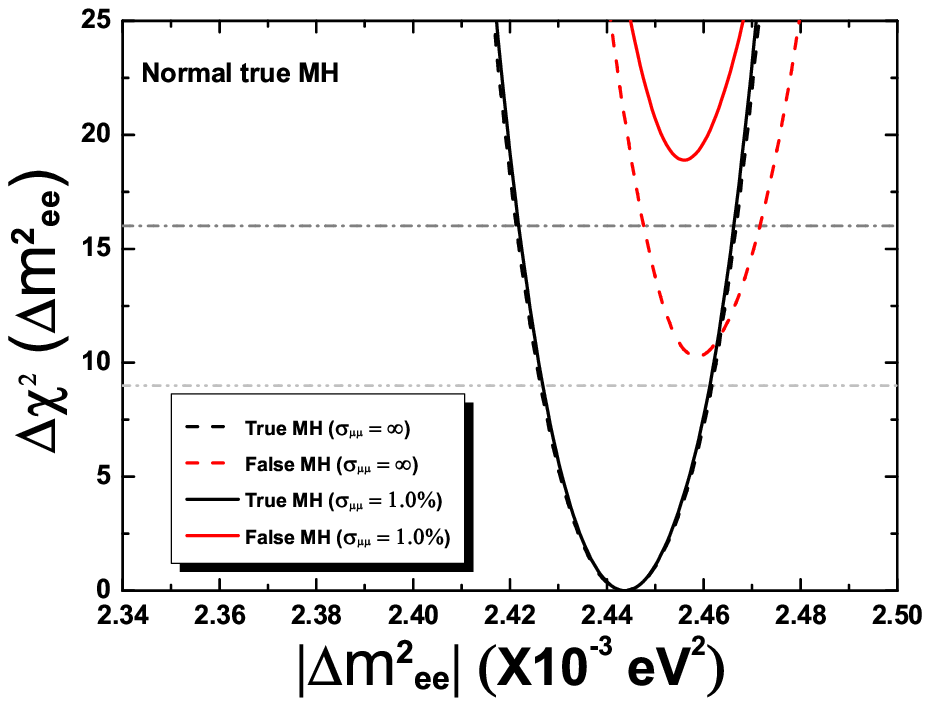}
&
\includegraphics*[bb=25 20 295 228, width=0.46\textwidth]{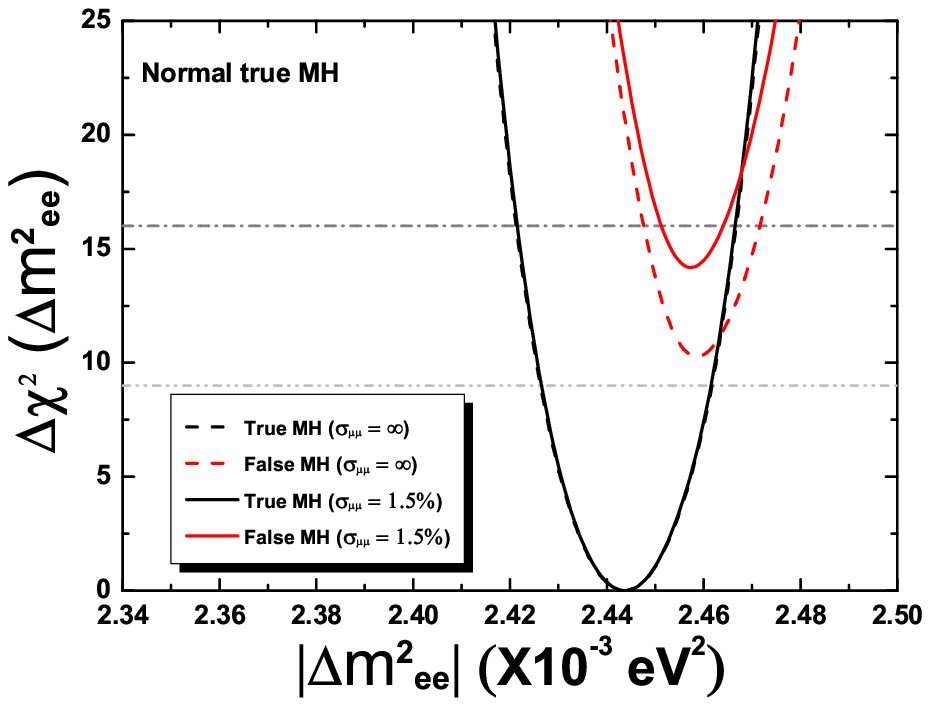}
\end{tabular}
\end{center}
\caption{The MH sensitivity of JUNO, where the comparison of black and red lines represents the discrimination power of the MH.
The solid and dashed lines are for the interference effects and the combined effects, respectively.
The uncertainties of the effective atmospheric mass-squared difference are taken as 1.0\% and 1.5\% for the left and
right panels, respectively.
\label{fig:MHsen}}
\end{figure}
The future prospective of the MH measurement at JUNO is shown in Fig.~\ref{fig:MHsen}, where a $3\sigma$ sensitivity is
achieved using the relative interference effects of reactor antineutrino oscillations~\cite{JUNO,Petcov1,Petcov2,Petcov3,dyb2t,dyb2e,sens}. 
The absolute spectral measurement taking account of both
interference effects and prior information of the effective atmospheric mass-squared difference can yield an improved sensitivity of 4$\sigma$. 
In our sensitivity studies, a 20-kton liquid scintillator detector with 3\% energy resolution and 6 years of data-taking time are assumed. In the left and
right panels of Fig.~\ref{fig:MHsen}, the uncertainties of the effective atmospheric mass-squared difference are taken as 1.0\% and 1.5\%,
and the corresponding MH sensitivity turns out to be 4.4$\sigma$ and 3.7$\sigma$, respectively.

\begin{table}[!htb]
\begin{center}
\begin{tabular}[c]{l|l|l|l|l|l} \hline\hline
  & Nominal &  + B2B (1\%)  & + BG &
  + EL (1\%) &  + NL (1\%) \\ \hline
 $\sin^2\theta_{12}$ & 0.54\%  & 0.60\% & 0.62\% & 0.64\% & 0.67\%  \\  \hline
 $\Delta m^2_{21}$ & 0.24\% & 0.27\%  & 0.29\% & 0.44\% & 0.59\% \\  \hline
 $|\Delta m^2_{ee}|$ & 0.27\% & 0.31\% & 0.31\% & 0.35\% & 0.44\%  \\
\hline\hline
\end{tabular}
\caption{\label{tab:prec} Precision of $\sin^2\theta_{12}$,
$\Delta m^2_{21}$ and $|\Delta m^2_{ee}|$ from the nominal setup to
those including additional systematic uncertainties. The systematics are added
one by one from left to right.}
\end{center}
\end{table}
As the prequisite to achieve the MH determination, the precision antineutrino spectral measurements from reactors provide an excellent opportunity to
precisely determine the oscillation parameters.
The JUNO experiment will be the first experiment to have an unprecedented precision measurement of $\sin^2\theta_{12}$, $\Delta m^2_{21}$ and $|\Delta m^2_{ee}|$
(defined as the linear combination of $\Delta m^2_{31}$ and $\Delta m^2_{32}$~~\cite{parke})
better than 1\%~\cite{JUNO,prec}. A systematic analysis of above-mentioned parameter measurements is summarized in Tab.~\ref{tab:prec}, where the nominal setup is defined as
the statistical uncertainty plus the reactor spectral uncertainty and detector uncertainty of 1\%. Additional systematic uncertainties from left to right
include the bin-to-bin (B2B) energy uncorrelated uncertainty, the background (BG), the energy linear scale (EL) and energy non-linear (NL) uncertainties.
In summary, for the precision measurements we can achieve the precision level of $0.5\%$$-$$0.7\%$ for the three oscillation parameters.
Moreover,  precision measurements of neutrinos are suitable to probe new physics beyond the standard three-neutrino paradigm,
including the hypotheses of light sterile neutrinos, nonstandard interactions, unitarity violation, and the Lorentz and CPT invariance violation.

As a large underground observatory, the JUNO experiment is also unique in the astrophysical neutrino measurements.
For a typical galactic supernova burst at a distance of 10 kpc, it will register about 5000 events from the inverse beta decay, 2000 events from
elastic neutrino-proton scattering, more than 300 events from neutrino-electron scattering,
as well as the charged current and neutral current interactions on the ${^{12}}{\rm C}$ nuclei.
the supernova neutrino observation at JUNO will provide a unique probe for both particle physics and astrophysics, including the absolute neutrino masses and the neutrino
mass hierarchy as well as identification of different phases of the supernova burst. Moreover, detection of the diffuse supernova neutrino background is also promising considering the detector size and background levels of JUNO.

There are many other aspects for the JUNO physics potential, including the observation of solar $^8$B and $^7$Be neutrinos, detection of geo-neutrinos, studies of atmospheric neutrino oscillations, indirect dark matter search using neutrinos, and observation of the proton decay in the $p \to K^+ + \overline\nu$ channel. All the physics studies are thoroughly summarized in Ref.~\cite{JUNO}.

\section{Experimental design}
\label{sec:physics}

\begin{figure}
\begin{center}
\begin{tabular}{c}
\includegraphics*[width=0.9\textwidth]{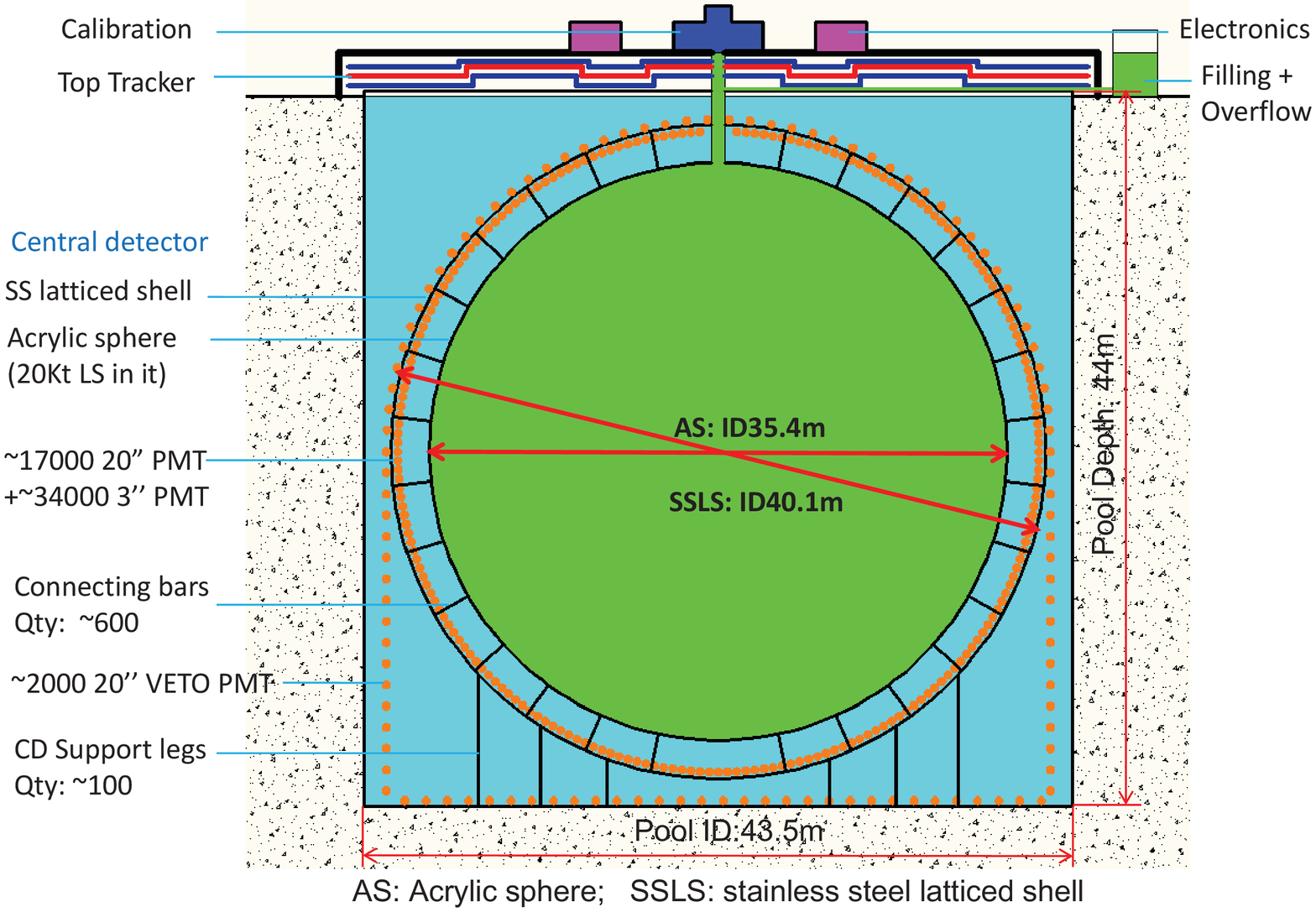}
\end{tabular}
\end{center}
\vspace{-0.8cm}
\caption{The nominal design of the JUNO detector system (see the context for details).
\label{fig:detector}}
\end{figure}
JUNO will deploy a LS detector of 20 kton in the underground experimental hall. The total overburden is $\sim$700 m rock. The baseline design of the JUNO detector system is
shown in Fig.~\ref{fig:detector}, which consists of the central detector (CD), the veto detectors and the systems for calibration, electronics, data acquisition
and detector control~\cite{JUNO2}.

The CD, which contains 20 kton LS, 17,000 20-inch PMTs and upto 34,000 3-inch PMTs, is designed to reach the energy resolution
of 3\%/$\sqrt{E_{\rm vis}}$ and the lifetime of over 20 years.
The CD~\cite{JUNO2} is an acrylic sphere of 35.4 m in diameter supported by stainless-steel structures,
where the acrylic sphere will be built from hundreds of acrylic sheets with a thickness of 12 cm using the bulk-polymerization technology.
The PMT assembly design including water sealing, implosion protection and installation on the detector is in progress, and a prototype to study
the performance of PMTs is under operation. The CD is filled with LS serving as the neutrino detection target.
The current LS recipe is linear alkyl benzene (LAB) as the solvent, with PPO as the scintillation fluor and bis-MSB as the wavelength-shifter.
The veto detector system consists of a Water Cherenkov (WC) detector and a top tracker using for the muon detection and muon induced background reduction.
The WC detector is a pool filled with purified water and instrumented with PMTs surrounding the CD.
The OPERA target tracker will be re-used as the top tracker to reconstruct the direction of cosmic muons.
The 20-inch PMTs are mounted on a spherical surface of $\sim$40 m in diameter, covering $\sim$75\% of the surface area of the sphere. The
total number of inward facing PMTs to detect scintillating photons generated in the CD is $\sim$17,000.
In addition, upto 34,000 3-inch PMTs will fill the gaps among the large PMTs in order to improve the energy and vertex resolution.

To achieve the primary goal of the MH determination, an unprecedented energy resolution of 3\%/$\sqrt{E_{\rm vis}}$ is a
critical parameter in the experimental design, which requires: a) the totoal photocathode coverage $\geq 75\%$, b) the PMT detection efficiency $\geq 27\%$
(defined as the quantum efficiency times the collection efficiency)~\cite{JUNO2,MCP,MCP2},
c) the LS attenuation length $\geq 20$ m at 430 nm, which is equivalent to an absorption length of 60 m with a Rayleigh scattering length of 30 m.
\footnote{The LAB Rayleigh scattering length is measured to be $28.2\pm1.0$ m at 430 nm~\cite{RS,RS2}.}

\section{Project status}

The JUNO project was approved by the Chinese Academy of Sciences under the Strategic Priority Research Program in 2013.
The international collaboration was formed in July of 2014, which includes over 300 collaborators and over 60 institutions from Asia, Europe and the Americas.
The groundbreaking ceremony at the experimental site was hold on January 10th, 2015. It would take around 3 years to finish the civil construction.
After the detector component production and the following detector installation and filling, JUNO is planed to start operation in 2020.

\section{Conclusion}

The primary goal of JUNO is to resolve the neutrino mass hierarchy and measure the oscillation parameters at the sub-percent level using reactor antineutrino oscillations.
Meanwhile, as a multipurpose neutrino experiment, it is also capable of measuring supernova burst neutrinos, the diffuse supernova neutrino background, geo-neutrinos,
solar neutrinos and atmospheric neutrinos. JUNO has received approval since 2013 and the civil construction has started from 2015.
The R$\&$D activities addressing the technical challenges are ongoing within the collaboration. JUNO plans to start data-taking in 2020.

\section*{Acknowledgments}

The author is grateful to Prof. Alexander I. Studenikin for kind invitation and warm hospitality in Moscow, where this wonderful conference was held.
This work was supported by the National Natural Science Foundation of China under grant Nos. 11135009 and 11305193, by the Strategic Priority Research Program of the Chinese
Academy of Sciences under Grant No.~XDA10010100, and the CAS Center for Excellence in Particle Physics (CCEPP).

\section*{References}

%
%
%
%
\end{document}